\documentclass{nime-alternate}

\begin{document}
%
\conferenceinfo{NIME'16,}{July 11-15, 2016, Griffith University, Brisbane, Australia.}

\title{Trends at NIME -- Reflections on Editing ``A NIME Reader''}

%
%
%
%
%

\numberofauthors{2} 
%
\author{
%
%
\alignauthor
Alexander Refsum Jensenius\\
       \affaddr{University of Oslo, Dept. of Musicology}\\
       \affaddr{Oslo, Norway}\\
       \email{a.r.jensenius@imv.uio.no}
\alignauthor
Michael J. Lyons\\
       \affaddr{Ritsumeikan University}\\
       \affaddr{Kyoto, Japan}\\
       \email{michael.lyons@gmail.com}
}
\date{January, 2016}

\maketitle
\begin{abstract}
This paper provides an overview of the process of editing the forthcoming anthology ``A NIME Reader---Fifteen years of New Interfaces for Musical Expression.'' The selection process is presented, and we reflect on some of the trends we have observed in re-discovering the collection of more than 1200 NIME papers published throughout the 15 yearlong history of the conference. An anthology is necessarily selective, and ours is no exception. As we present in this paper, the aim has been to  represent the wide range of artistic, scientific, and technological approaches that characterize the NIME conference. The anthology also includes critical discourse, and through acknowledgment of the strengths and weaknesses of the NIME community, we propose activities that could further diversify and strengthen the field.
\end{abstract}


%
%

\acmclassification1998{H.5.5 [Information Interfaces and Presentation] Sound and Music Computing}

\section{Introduction}

Over its 15 years of existence, NIME has become an important conference series. Each year NIME is the meeting point of researchers, developers, designers, and artists from all over the world. Even though participants come from widely different backgrounds, they share a mutual interest in groundbreaking music and technology. 

NIME has grown from a small workshop at CHI in 2001 \cite{Poupyrev:2001a}, to a large-scale conference with up to 500 submissions per year. In fact, more than 1200 papers have been published through the conference so far, and, staying true to the open and inclusive atmosphere of the community, all of the papers are freely available online.\footnote{http://www.nime.org/archive/} The archive is great if you know what to look for, but it has grown to a size that is difficult to handle for newcomers to the field. Even for long-timers and occasional visitors, it is difficult to get an overview of the history and development of the community. 

As frequent participants at NIME, but also as co-founder (Lyons) and current chair of the Steering Committee (Jensenius), we are happy to see that a growing number of papers focus on historical, theoretical and reflective studies of the NIME community itself. We believe this is a healthy next step for the community to move on to become a field in its own right. As the level of meta-studies started to grow, we began to see the potential for a collection of articles to broadly represent the conference series. This thought has now materialized in the anthology ``A NIME Reader---Fifteen years of New Interfaces for Musical Expression,'' to be published by Springer Verlag in 2016  \cite{jensenius_nime_2016}. This anthology, which will refer to as \textit{Reader} in the rest of this paper, consists of 30 selected papers from the conference series, and with two new commentaries written for each paper. 

The Reader is intended for everyone interested in the field, ranging from newcomers who just want to get an overview of some of the recent advances in music technology, including important interfaces, technologies and artistic outcomes, to researchers interested in knowing about the meta-discussions and reflections in the field. The selection of papers reflects both the depth and the width of the publication archive, and the commentaries that follow each paper have added to the value of the original papers while at the same time help bring some important underlying discussions alive throughout the book. 

The current paper is offered with several intentions. First of all we want to share openly with the community the selection process we underwent when choosing the 30 articles for the book as well as our experience going through the entire NIME archive as part of this process. We also want to highlight some of the insights gained as part of the selection, review, and peer commentary processes, some of which are well known, but some of which took us by surprise. From this we propose some actions that respond to issues and insights resulting from the editing project. Finally, we want to encourage others to take up meta-study and anthology projects of their own in order to  ``digest'' the (already massive) growing literature of NIME.

\section{The Selection Process}

By nature, an anthology is a limited selection of chapters, and for this project we settled on a limit of 30 articles from the NIME archive. This number was arrived at by estimating the page count of the resulting volume, and it also fitted well with the idea of including approximately two published items for each year of the NIME conference up to and including the most recent edition in 2015. 

When we set out to create this volume, neither of us doubted it would be a challenging task. After all, how does one select 30 papers from more than 1200 items that have been published in the NIME proceedings, as of 2015? How could we possibly fairly represent the energetic and creative output in the field of musical interface research, post-NIME? From the outset, the NIME community has intentionally prioritized a diversity of research styles and approaches. The conference has also striven to offer an environment that can attract the participation not only of researchers working in an academic or institutional context, but also independent artists, researchers and inventors. Moreover, each of us has our own subjective interests and tastes as to what we consider significant prior work.

We started the selection process by each creating an initial list of 50 or so articles perceived as ``influential'' by the community. This was done by identifying the most-cited NIME papers in the Google Scholar index. There are other bibliometric sources, but we found it important to use an openly available index. Naturally, older papers tend to have more citations than new ones, so we also considered a measure of the number of citations-per-year, by dividing the total citation count by the number of years since publication. Both measures were used to create our ``objective,'' or more accurately ``multi-subjective,'' initial list of works that have enjoyed impact.

From this first list of well-cited works, which at least to some extent tells us that other researchers have shown interest in a particular paper, we systematically examined the content of each one. We each separately created our lists of papers to include. In many cases, we agreed fully, but for others discussion, and sometimes multiple discussions, were needed before we reached agreement. At the same time, we found that the initial list had gaps in the coverage of some topics and research approaches. For example, work that is primarily artistic in nature is usually not as highly cited as a report on an important technological advance. This does not imply that artistic research papers are less excellent, important, or influential on the direction of subsequent work, but there is just a lower number of active researchers with a primarily artistic focus. 
This led each of us to propose some works that have not been cited as highly as the others, but which we believe represent some of the diversity of the NIME corpus. Much discussion was also needed to reach agreement on which of these were to be included in the final list.

Needless to say, we have not been able to include all of the papers that we believe deserve a place in a NIME anthology. The final selection reflects a compromise to keep within the limit of just 30 articles. This is also why it is consciously titled ``\emph{A} NIME Reader'' and not ``\emph{The} NIME Reader.'' We hope that there will be other anthology projects drawing on the extensive NIME proceedings in the future. That is also why we towards the end of this article suggest projects that may be of interest for future work.

The final selection of papers is presented in Table~\ref{tab:papers}. The list covers the entire history of the NIME conferences, although we for different reasons were not able to include exactly two papers for each year. 

\begin{table*}[tb]
       \caption{List of NIME papers included in the anthology}
       \label{tab:papers}
       \centering

       \begin{tabular}{l|p{0.4\textwidth}|p{0.4\textwidth}}

\textbf{Year/Item} & \textbf{Title} & \textbf{Author(s)}\\
       \hline
2001a & Principles for Designing Computer Music Controllers & Cook\\
2001b & Problems and Prospects for Intimate Musical Control of Computers & Wessel \& Wright\\
2002a & The importance of parameter mapping in electronic instrument design & Hunt, Wanderley \& Paradis\\
2002b & Multimodal Interaction in Music Using the Electromyogram and Relative Position Sensing & Tanaka \& Knapp\\
2002c & The Plank: Designing a Simple Haptic Controller & Verplank, Gurevich \& Mathews\\
2003a & Contexts of Collaborative Musical Experiences & Blaine \& Fels\\
2003b & Sonigraphical Instruments: From FMOL to the reacTable*  & Jord\`{a}\\
2003c & Designing, Playing, and Performing with a Vision-based Mouth Interface & Lyons, Haehnel \& Tetsutani\\
2003d & OpenSound Control: State of the Art 2003 & Wright, Freed \& Momeni\\
2004a & The Electronic Sitar Controller & Kapur, Lazier, Davidson, Wilson \& Cook\\
2004b & PebbleBox and CrumbleBag: Tactile Interfaces for Granular Synthesis & O'Modhrain \& Essl\\
2004c & Toward a Generalized Friction Controller: From the Bowed String to Unusual Musical Instruments & Serafin \& Young\\
2004d & On-the-fly Programming: Using Code as an Expressive Musical Instrument & Wang \& Cook\\
2005a & Towards a Dimension Space for Musical Devices & Birnbaum, Fiebrink, Malloch \& Wanderley\\
2005b & The Overtone Violin &  Overholt\\
2006a & Sensemble: A Wireless, Compact, Multi-User Sensor System for Interactive Dance & Aylward \& Paradiso\\
2006b & Mobile Music Technology: Report on an Emerging Community & Gaye, Holmquist,  Behrendt \& Tanaka\\
2007a & Wireless Sensor Interface and Gesture-Follower for Music Pedagogy & Bevilacqua, Gu\'{e}dy, Schnell, Fl\'{e}ty \& Leroy\\
2007b & Don't Forget the Laptop: Using Native Input Capabilities for Expressive Musical Control & Fiebrink, Wang \& Cook\\
2007c & Expression and its Discontents: Toward an Ecology of Musical Creation & Gurevich \& Trevi\~{n}o\\
2007d & The Acoustic, the Digital and the Body: A Survey on Musical Instruments & Magnusson \& Hurtado\\
2008 & Eight Years of Practice on the Hyper-Flute: Technological and Musical Perspectives & Palacio-Quintin\\
2009 & A History of Hemispherical Speakers at Princeton, Plus a DIY Guide & Smallwood, Cook, Trueman \& McIntyre\\
2011 & Satellite CCRMA: A Musical Interaction and Sound Synthesis Platform & Berdahl \& Ju\\
2012a & The Fingerphone: a Case Study of Sustainable Instrument Redesign & Freed\\
2012b & To be inside someone else's dream: Music for Sleeping \& Waking Minds & Ouzounian, Knapp, Lyon \& DuBois\\
2012c & TouchKeys: Capacitive Multi-Touch Sensing on a Physical Keyboard & McPherson\\
2013 & The Web Browser As Synthesizer And Interface & Roberts, Wakefield \& Wright\\
2014 & To Gesture or Not? An Analysis of Terminology in NIME Proceedings 2001--2013 & Jensenius\\
2015 & Fourteen Years of NIME: The Value and Meaning of `Community' in Interactive Music Research & Marquez-Borbon \& Stapleton\\
\hline

\end{tabular}
\end{table*}

\section{Peer-Commentary Process}

An important feature of the Reader is the inclusion of not only articles selected to introduce and represent NIME research, but commentaries on those articles written both by the authors themselves as well as by other knowledgeable participants in the NIME community (henceforth ``experts'' or ``peers.'') Each article is accompanied by one commentary by one or several of the original authors and one expert commentary. Moreover, authors and peers had a chance to exchange feedback on each other's commentary. The feedback ranged from simple corrections to enthusiastic discussions of philosophical issues and resulted in revision of the commentary texts in the majority of cases. We, as editors, took part in this process both by facilitating and moderating the exchanges, as well as in providing suggestions and commentaries ourselves. We found that the peer-commentary procedure, inspired by Stevan Harnad's thinking and writing on the subject \cite{Harnad:2000}, really brought our project to life: new ideas and insights in the peer commentaries and in the exchanges between authors and experts showed how the selected articles continue to live, function, and contribute growth in the research community.

Note however, that the peer-commentary process is not the same thing as peer-review, that is, the use of single or double blind peer comments aimed at improving the papers. All of the included articles have already undergone a peer-review process in order to be published at NIME, so we agreed that the papers should not be updated or expanded, except for updating/removing broken URLs and making formatting adjustments. We also found that carrying out peer-review and selection of the commentaries would be overkill, as the commentaries were mainly seen as a way to create discussion and engagement. 

Readers may wonder about the conditions of the commentary process: how were the experts selected, and how was the commentary elicited?  As for the selection of articles, the choice of commentators was negotiated between the two editors, through creating and discussing initial lists. Many, but not all, of the commentators selected were perceived to be frequent ``users '' or supporters of a given article, for having cited it frequently. In other cases, commentators were selected to increase the inclusiveness (we admit this is difficult to objectively define) of the participants in the anthology, or to involve regular conference participants who were not otherwise involved in the project.  Here again, limitations of the scale of the project meant that we were not able to involve everyone who came to mind and we recognize that many who have much to contribute were not included as commentators. By the same token, some we initially approached were not able to join as commentators for various reasons. 

\section{Observed Trends}

The lengthy process of selecting articles from the extensive NIME conference proceedings as well as through organizing and moderating the peer-commentary activity, afforded an opportunity to reflect on trends that the NIME community has experienced in the past decade and a half. Some of these trends are well-established, others are still developing, while others again are still in their infancy. While detailed description and discussion of these is beyond the scope of the current paper, our intention in this section is to outline some of the major trends we have observed. In the following, each point is supported by references to articles from the Reader, as indexed in Table~\ref{tab:papers}.

\subsection{Toy to Instrument}
The NIME community has greatly benefited from the tremendous progress in human interface technology over the past two decades. Not surprisingly, many of the articles in the Reader introduce, or make use of, improved sensor technologies (e.g. 2002c, 2003b,c; 2004a,c; 2005b, 2006a, 2007a, 2012c). By the same token, interface micro-controllers for data acquisition and device control have improved in every aspect and become easier to use, while higher resolution, faster, and more flexible communication protocols were developed, standardized, and are now widely in use (e.g. 2003, 2011). Instrument designs have made more effective and engaging use of the richer, cleaner, high bandwidth data streams available, with more complex motion-to-synthesis parameter mappings (e.g. 2001b; 2002a; 2003b). 

More generally, the approaches to the design of new musical interfaces have become better informed and employ increasingly sophisticated and theoretically better grounded design strategies (e.g. 2001a; 2003b; 2004b; 2005a; 2007c, d; 2012a; 2014). Indeed, we now perceive this to be the longest and strongest ``trend'' or thread in the NIME conference and it shows no sign of abating. More strictly formal approaches to instrument designs have, in recent years, employed rigorous machine learning techniques (e.g. 2007a). We acknowledge this as a highly significant and growing trend at NIME, but have chosen consciously not to attempt to cover this topic fully in the Reader, but to leave it as an ``advanced topic'' for more specialized reviews. Similarly, progress in computing hardware and software allows increasingly powerful approaches to sound synthesis (e.g. 2004d; 2009, 2011).

\subsection{Buttons to Embodiment}
A key motivation behind human-interface research of the past two or more decades has been to move away from the keyboard-mouse-windows concept of HCI towards more fluid, full-body interaction. NIME is no exception, and indeed the rapid expansion and progress in this area was one of the factors to which the conference owes its existence. Accordingly, research into motion sensing and music making continues to form a significant thread (e.g. 2003b,c; 2004a; 2005b; 2007a). Even more intimately, new musical interfaces have striven to use signals inside the body, through biosensing technologies (e.g. 2002b; 2012b), or to capture motion of the whole body (e.g. 2003c; 2006a; 2014).

\subsection{Individual to Community, Stage to Street}
One emerging trend in NIME research has been the move from individual performance to interfaces that put the focus on collaborative creation (e.g. 2003a; 2006a). Likewise, some researchers have explored technologies and approaches towards supporting group music-making via laptop orchestras (2007b; 2009), web-based performance (2013), and made the technology development process itself public and collaborative via live coding activities (2004d). Another strong thread is the use of mobile technology to create music anywhere, anytime (2006b). This sub-community of NIME emerged so strongly that it has also spawned independent conferences and workshops, a process that has now also begun for the live-coding interest group.

\subsection{``Hacking'' to ``Professionalism''}
Early NIME conferences had an aspect of ``Woodstock-like'' gatherings of tinkerers, hackers, and makers. However, the music technology research community has grown and matured over the years. While a healthy respect for adhoc, improvised approaches persists, we also see individuals and groups engage in more long-term and structured development work. This work is often focused on development as \emph{process}, with an acknowledgment of both formal and informal evaluation of the interfaces as an important part of this process. Part of such evaluation can be that of a larger commitment to continued performance with new instruments (e.g. 2008) in contrast to the early preponderance of ``demo and die'' interfaces. 

As part of the increased ``professionalism'' of the field, we are happy to see that new interfaces have often inspired crowd-funding campaigns, which again have lead to the funding of successful companies. A prominent example of successful commercialization is Smule, which offers music-making apps for mobile platforms. Similarly, well-supported open source software and hardware projects have flourished (e.g. 2011, 2012c) and attracted substantial participation. Introductory courses \cite{Fels:2009,jensenius_action-sound_2013,Lyons:2014,Tzanetakis:2013} and outreach activities  \cite{Bevilacqua:2013} have also increased in frequency.

With the increasing maturity of the field, we see that more papers are also devoted to topics outside the ``doing'' of NIME, including: pedagogy, history and various types of reflection on the technologies, the artistic outcomes and on the community itself.

\subsection{The NIME Community}
The most recent article to be included in the Reader, from NIME 2015, comes from Adnan Marquez-Borbon and Paul Stapleton \cite{marquez-borbon_fourteen_2015}. They offer a critical analysis of the NIME community, concluding that it more closely resembles a loosely organized \emph{community of interest} than a well-focused \emph{community of practice}. Furthermore, they observe that a lack of a mechanism for judging what can be considered a ``good'' NIME performance may inhibit progress in the field. Their commentary to their own article goes further to suggest that a perceived lack of ``coherent vision and critical reflection'' may be marginalizing or excluding those who are not part of the ``dominant social norm.''

Critical discussions of the NIME conferences (and community) have, of course, taken place for many years---in informal discussions among members, in the conference Steering Committee, and also at the annual ``town hall'' meetings at the last day of the conference. However, the appearance of formal critical discourse as an official component of the academic program, such as that of Marquez-Borbon and Stapleton, marks maturity of the conference and an increasing self-awareness of NIME. Such considerations suggest that it may be time to engage in research into various aspects of the NIME community. 

Many approaches can be envisaged here, from scientometric analyses of the NIME proceedings, such as Jensenius' study of the terminology of gesture \cite{Jensenius:2014}. It can also be interesting to carry out sociological or ethnographic studies, such as that of Born's study of IRCAM \cite{born_rationalizing_1995}. Born and Devine have also recently studied the gender (im)balance in higher music technology education in Britain \cite{born_music_2015}, a topic which is also highly relevant for the still male-dominated NIME community. To start with, however, a modest proposal that arose during the peer commentary process on the Marquez-Borbon and Stapleton article, is that it would be valuable to survey the members of the community itself about their experiences to date and expectations for how NIME should further develop as a community.

\section{Future Directions: How can NIME Continue to be Relevant?}

In closing we would like to raise the broader question of the continuing relevance of the NIME conference. With the advent of a number of other conferences, festivals, and venues for presentation of research and artistic practice related to music technology, we recognize a need for the community that has emerged around the conference to reflect on what has been accomplished over the past decade and a half, and to consider how to move forward:

\begin{itemize}
\item What aspects of this community's identity are strongest? 
\item What shortcomings and imbalances need to be addressed? 
\item What interests and viewpoints have been neglected and how can positive action be taken to cover these? 
\item How can we organize what has been accomplished in such a form that it is accessible and useful for new generations of researchers and artists? 
\item How can we promote the activities of our community more widely and support long term activities such as project repositories, development platforms, and entrepreneurship? 
\end{itemize}

These are questions for everyone who participates in NIME and we hope that our anthology project and this article will encourage everyone to become active in the process of reflecting on what has been achieved and contributing to deciding the directions the community will take in the future.

\subsection{``Digesting'' the past}

One way in which to focus the development of the community itself is through projects that ``digest'' past and ongoing work within the field. Our anthology is one example of such a project. We would like to suggest projects that go beyond our Reader. For example, it would be interesting to create a platform that allows ongoing, open (but moderated), peer-commentary of articles selected into the NIME proceedings. This would allow dialogue on all works to continue openly throughout the year, after the intense few days of the annual conference have ended. Participation could be made voluntary, but we can imagine that this would be attractive for most researchers if it could be made to function smoothly. Such ongoing dialogue on a project could help with: 

\begin{itemize}
\item Nurturing ideas: many NIME papers are fairly terse and have only room to present one (or a few) core ideas of a larger picture. It would be useful to create a space in which ideas can be expanded, generating new insights, suggesting new research directions, and supporting community-building. 
\item Discovering pre-existing ideas and works: the NIME publication archive is already an incredibly rich resource for discovering neglected ideas and works, but augmenting this with peer-commentary would serve to make the material more transparent and accessible.
\item Educating new NIMErs: as a still relatively new community, with few books and the lack of a long history or consensus, newcomers to the field are often bewildered as to where to start looking for information. There is clearly a need to develop material that is useful in education but also for researchers in other fields visiting the NIME community. Such material could be a useful spin-off of an ongoing peer-commentary platform.
\end{itemize}

Another idea is to create an active, living repository of musical interface designs, which allow creators to share code, hardware specifications, data, and other material associated with a given project. This would work towards treating the ``demo and die'' syndrome, and serve as a valuable resource also for composers and performers who would like to leverage already developed technology for artistic purposes.



\subsection{Anthologies to Come} 

It is important to stress that even though we have made a collection now, the aim has never been to create the ultimate NIME paper selection, or a definitive ``canon'' of relevant work. We would therefore like to provide some concrete suggestions for others interested in conducting a review or creating an anthology: 

\begin{itemize}
\item Anthologies on more specialized topics, for example on mobile or web technologies, machine learning, or on the importance of tactility, artistic use, performance considerations, etc.
\item Anthologies based on a particular theoretic thrust, such as gesture, liveness, community development, pedagogy etc. 
\item Anthologies based on works that have not received much attention, for example, least cited works, odd technologies, or comical works?
\end{itemize}

In conclusion, we hope that our collection will be the first of many. After all, the NIME proceedings archive is a gold mine of good ideas, and a history of experience and knowledge gained through the dedicated work of many talented researchers and artists. These should not be forgotten but rather continue to be used in different ways. 

%
\bibliographystyle{abbrv}
\bibliography{bibliography}  
%
%


\end{document}